# The Hyper-Cortex of Human Collective-Intelligence Systems

Marko A. Rodriguez

Computer Science Department
University of California at Santa Cruz
and
Center for Evolution, Complexity, and Cognition
Vrije Universiteit Brussel, Belgium

http://www.soe.ucsc.edu/~okram
okram@soe.ucsc.edu

## Abstract

Individual-intelligence research, from a neurological perspective, describes the cortex as a medium for performing conceptual abstraction and specification. This idea has been used to explain how motor-cortex regions responsible for different behavioral modalities such as writing and speaking can express the same general concept represented in the cortex. For example, the concept of a dog, abstractly represented in the higher-layers of the cortex, can either be written or spoken about depending on the context. Abstract models in the higher-layers propagate activation patterns down the cortical hierarchy to the desired region of the motor-cortex for worldly implementation. In this paper, the individual-intelligence framework is expanded to incorporate collective-intelligence within a hyper-cortical construct. This hyper-cortex is a multi-layered network used to represent abstract collective concepts. This collective-intelligence framework plays an important role in understanding how collective-intelligence systems can be engineered to handle collective problem-solving. To conclude the paper, five common problems in the scientific community are solved using an artificial hyper-cortex generated from digital-library metadata.

## 1. Introduction

Research published by Jeff Hawkins and Sandra Blakeslee [1] has provided the foundation for many of the ideas in this paper. Hawkins and Blakeslee define the human cortex as a predictive system that pattern-matches the current sensory experience to an analogous concept in memory. Multiple subjectively similar events are mapped to a single abstract model. These lossy models are called *invariant representations*. Invariant representations in the cortex may be associated with a specific learnt behavior. In such cases, if the associated behavior proved successful in the past, then the individual executes the behavior relative to current context. In other instances, the invariant model may trigger a continuous cascade of associated memories—other invariant representations. In such situations, the individual performs multiple internal transformations of the input signal in the process of conceptual thinking. In short, the cortex is a memory structure that associates present stimuli with previously held concepts about the world. Pattern-matching will be shown to be the cortical function required for individual problem-solving.

The paper then extends the individual cortex model to encapsulate a socially-embedded *hyper-cortex*[1] residing outside the individual, existing within the collective. The layers of a hyper-cortex represent information which is unable to be fully contained within any single individual. These hyper-cortical layers are created by the aggregate interactions of multiple individuals over time. In order to represent abstract collective concepts, a hierarchy of layers can cluster similar lower layer regions into subjectively related ideas. During specific problem-solving instances, an individual who is unable to locate an appropriate invariant solution within their cortex can use the collective's hyper-cortex to derive a solution. In effect, a hyper-cortex functions as a pattern-matching infrastructure—mapping user realized problems to collectively-derived solutions.

---

[1] Discussed by Pierre Lévy [2] but originally coined by Pierre Teilhard de Chardin

Outlining the paper's structure, the initial section (Section 2) explains the general individual-intelligence framework described by Hawkins and Blakeslee [1] and further formalized more concretely by other researchers in the neurosciences. Section 3 will extend this framework into the domain of collective-intelligence. In Sections 4, an artificial hyper-cortex is generated from digital-library metadata, and in Section 5, the digital-library hyper-cortex is used to solve five problems commonly encountered within the scientific community.

1. *locating references for an initial idea*: the hyper-cortex correlates a collection of related manuscripts with an individual's rough and unformulated idea.

2. *finding potential collaborators for an idea*: given a model of the specific domain of the individual's initial idea, the hyper-cortex will suggest fit collaborators.

3. *determining a journal for paper submission*: once the collaborating team has formalized the idea into a written manuscript, a journal suitable for the paper will be recommended by the hyper-cortex.

4. *locating peer-reviewers to review a paper*: journal editors can use the hyper-cortex to find qualified referees to review the submitted manuscript.

5. *distribute the accepted paper to appropriate members in the community*: the hyper-cortex creates a plausible mapping between the newly published paper and potentially interested scientists in the field.

## 2. Individual-Intelligence in the Cortex

Before delving into this paper's collective-intelligence framework, discussion of the human cortex's role in individual-intelligence is necessary. Individual problem-solving will be defined as the cortical act of matching a given problem's context to potential solutions previously existing in memory (Subsections 2a,2b&2c). The section concludes with a real-world example of an individual solving a problematic and unexpected event—an event that can't be represented by the cortex—by utilizing another individual's memory of the solution (Subsection 2d) .

### 2a. Mapping Sensory Information to Experiential Memories

Any individual's sensory input is represented within their cortex as a series of activation patterns situated within space and/or time. Propagating sensory patterns associate themselves with an individual's stored subjective categorization of the external world. These internal categories are called *invariant representations* and are formed as the individual's cortex organizes the countless, different but relatable everyday experiences into abstract generalizations [3]. Invariant models possess a one-to-many compression ratio with their continuously fluxing sensory equivalents. That is, many experiences are mapped by the individual's cortex to one invariant representation. The term 'invariant' expresses the idea that, even under various transformations of the input signal (i.e. rotations, translations, scalings, etc.) the same abstract model can represent all transformations of the signal [4]. The invariant representation's modeling ability is impervious to slight variations of the input signal, and is therefore an approximation of many related, yet different occurrences. Due to this compression functionality, invariant representations form the foundation of conceptual thinking, the process of generalization and abstraction.

A more complex concept, one relying on multiple sensory modalities, is elicited when the multiple sensory input signals ascend the cortical hierarchy and converge on one single invariant model. This is how the smell, sight, and sound derived from a person can unite to activate the individual's pre-existing concept of a well known friend [5]. Furthermore, even without a visual signal from a friend's physical presence, an auditory signal—e.g. a friend's voice heard from behind—will prime the individual to expect seeing their friend upon turning around. This example illustrates that the human cortex can use subsets of an invariant model to predict future events—anticipation [6]. Over time, as more sensory information related to the invariant model of the friend is made available, the individual will further affirm their subjective belief that the friend co-exists with them in physical reality.

Invariant representations, stimulated by their analogous sensory information, can trigger a cascade of memories, which lead to predictions and behaviors. Within this framework, it is shown that predictions and behaviors are one in the same phenomenon [1]. Predictions and behaviors are activation patterns of invariant models, descending down the hierarchy to either; prime sensory cells to anticipate the next instant's experience, or elicit certain behaviors attempting to increase the predictability of the individual's environment. With any sensory input the individual's interpretation of the signal will be unavoidably associated with their previously held invariant concepts about their world. An individual thinks and acts with a continuous and innate reliance on their internal memories. If and when expectations—derived from memory—are not met, invariant models will be altered. Thereby the human cortex organizes a diverse and seemingly unpredictable world into a more consistent subjective experience. Invariant representations change throughout the individual's life as unexpected experiences adversely associate with inappropriate internal models [4]. A constant process of mapping the present and the past into predictable and more palatable cortical models of reality is at the root of intelligent problem-solving.

## 2b. Hierarchical and Lateral Projections within the Cortex

The hierarchical structure of the human cortex is composed of interconnecting layers of neural tissue. Neurons on the higher levels of the cortex's structural hierarchy have greater invariance to sensory signal fluctuations than do those located on lower levels [7]. A cluster of lower-layered neurons, each of which project to a single higher-layer neuron, is called the invariant neuron's *receptive field* (Figure 1). Only the lower-level layers are responsive to small deviations of the sensory signal (i.e. light contrast, lines, and basic shapes). When considering a conceptually stable object experienced in the world, the activation pattern in the receptive field of the object's invariant model varies significantly through time. The invariant model of that object, on the other hand, maintains a constant firing pattern. Therefore, an object can move through the individual's visual scene and still be recognized as the same object.

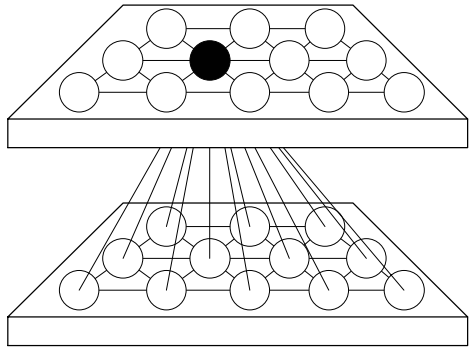

*Figure 1: the receptive field of the abstract neuron (black) is a cluster of connected lower-layer neurons*

Within the layers of the cortex, neurons are also connected with horizontal projections. Horizontal projections within a cortical layer can produce a broadening or contraction of a sensory signal's potential categorization [8]. Partial excitation of a receptive field can spread over the entire receptive field, causing the invariant neuron of that receptive field to fire. For example, an object can still be recognized even if it is partially hidden from view. On the other hand, inhibitory projections to remote receptive fields can dampen the probability of the remote receptive field stimulating its respective invariant representation [9]. When a receptive field inhibits the activation of a neighboring receptive field, the sensory experience will be contained to one abstract interpretation. Lateral inhibition has been studied extensively in the visual system [10]. Without lateral inhibition the same sensory experience would be mapped to multiple unrelated concepts. Without lateral excitation sensory experiences would seem too varied and inconsistent to be grouped above and beyond the lower-layer's excitation pattern. A fine balance between these two algorithms provides the human with a chaotic-stable experience of reality.

## 2c. Problem-Solving Sensory Perturbations

A *problem*, in its most basic form, is any sensory input—or *perturbation*. When sensory information enters through the individual's sensory modalities, the cortex begins to categorize the information into concepts of higher and higher order. A cascade of action potentials throughout the cortex causes the individual to either contain the signal internally as a rationalization in thought, or externally as a behavior that alters the external environment. This resolution of sensory perturbations is called *problem-solving*. Signal resolution can happen through priming or through action. In priming, the current sensory pattern can cause an invariant model to prime its receptive field to expect a particular pattern at the next instant. If the expectation is met, then the receptive field is fully excited and thus the invariant model neuron is activated again. Other times, expectations are not met and the primed cortical region does not receive enough excitation from the sensory signal to elicit the experience of the concept. In this case, the individual has a parallel experience of reality—one is their *memory-based expectation* and the other is their *sensory-based experience*. When a discrepancy between expectation and experience occurs, a larger problem for the individual exists. In such cases, the sensory signal continues to elicit activations laterally and vertically through the cortex until an appropriate model of the signal is located. For example, a cortex that has been habituated to a constant sensory signal through overexposure is unperturbed by continuous exposure to that same signal. The signal causes a higher-level invariant representation to continuously prime the low-level cortical neurons to expect that same signal at the next time step. If that expectation is met, then the signal is easily resolved as an instance of the invariant representation. In this case, the future is constantly being accurately predicted and therefore the problem-solving environment is simple—the individual contains the appropriate internal model to resolve the sensory perturbation. If at any time in the future the overexposed signal is altered in such a way that the primed receptive field of the invariant representation isn't activated, then the cortex's expectation is not met. In not meeting the prediction, a rationalization of the altered signal's properties must be deduced. Sometimes the signal can be resolved in the head and other times it requires the active manipulation of the environment—which in turn begins the process again. The human cortex is constantly problem-solving by changing its internal structure and its external world to ensure accurate predictions through time. The human cortex is constantly reaching for equilibrium by through constant adaptation of its internal structure in order to accurately represent its external world.

## 2d. Individual-Intelligence Problem-Solving

Let's apply this general framework to a simple individual-intelligence problem-solving scenario. Suppose an individual is working in his or her office. As the individual scans the room, each item in the office streams through the individual's visual system via a spatial-temporal pattern of activation across their retina (Figure 2a-P). If the human's neural-network is well adapted to the office then the sensory pattern is easily mapped to an invariant representation—the concept of their room and the standard items it contains. Nothing is out of the ordinary since the signal matches the primed invariant model (Figure 2a-S). In this example, the activation pattern further habituates the neural-network to the external world and stimulates no active behavior in the world. The individual's predictions of the world are true and the individual is unconscious of any problem.

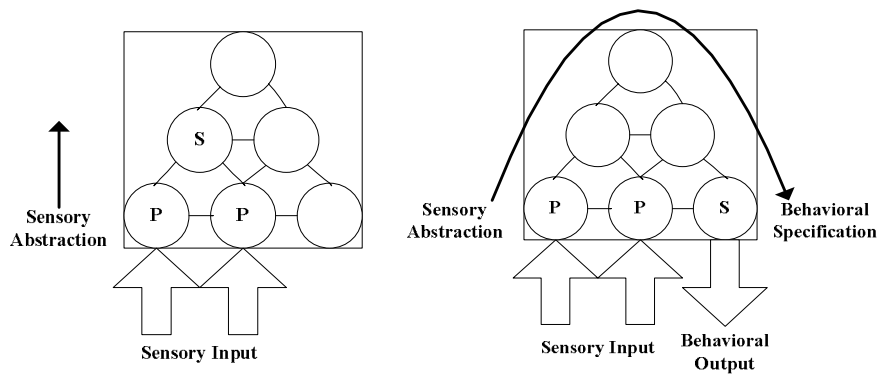

*Figure 2a&b: internal and external problem-solving resolution. (P) problem-model and (S) for solution-model*

Now if at some point in time a new item is placed into this individual's office then the input pattern is not so effectively mapped to the invariant representation of the office (Figure 2b-P). Though some aspects of the office model will fire (i.e. that it's the same office, same chair, same desk, etc.), the other aspects of the signal concerning the new item will have to find resolution elsewhere in the cortex. To handle this perturbation, the cortex performs a series of transformation on the unresolved portion of the input signal until an appropriate understanding of the new items existence is realized. For this example, a potential abstract solution is that someone in the house must have placed the new item in the room. In gross simplification, this abstract solution propagates an activation pattern down the cortical hierarchy to the individual's motor-cortex. The activation of the motor-cortex prompts the individual to leave the office and ask others in the house about the new item (Figure 2b-S). The behavior output of this complex problem-solving process is the physical act of questioning everyone for a solution. This is the beginning of collective intelligence.

Many of the ideas outlined in this section lack strong boundaries between them. For instance, memories, invariant representations, internal models, concepts, and habituated cortical-regions are all essentially the same idea. When discussing the diversity of expression allowed by the cortex through its fundamental properties of a layered network, specific ideas are difficult to contain.

# 3. Collective-Intelligence in the Hyper-Cortex

The individual-intelligence framework presented in the previous section was a mix of cybernetic concepts and ideas outlined by Hawkins and Blakeslee. The depth of their book has not been done justice as only those ideas pertaining to the goal of this paper have been outlined. It is recommended that the reader read their book for a greater understanding of this simple yet effective model of human intelligence. This section will now generalize the individual-intelligence model to represent collective-intelligence within a hyper-cortex. This hyper-cortex maintains the highest layers of a societal problem-solving hierarchy.

### 3a. Memory in a Hyper-Cortex

As stated in the previous section, the cortex can be generically described as a memory system. Each experience contributes to the adaptation of the cortical structure—habituating the cortical structure to accurately associate its sensory experiences with stable container concepts. These concepts allow for the many-to-one relationship seen when the ever flux experience of the external world is mapped to a conceptually consistent internal explanation. It is only through habituation (learning) and analogy (thinking) that this stability emerges. External to the individual, within the domain of the collective, there also exists such networks of habituation and analogy. The learning web system formulated in [11] is a collectively created network of associated concepts. As individuals traverse the word-network, linking from concept to concept, they strengthen old associations and consequently weaken others. The learning web is a network which continuously adapts its structure to learn the collective preference of its users. This collectively created word-network has been called a *collective mental-map* [11]. What other such maps exist in the world today? The World Wide Web is a network of related web pages, tied by their associations within domains subjectively determined by their authors. Unlike the cortical structure of the human brain, the word-network and the World Wide Web are both flat hierarchies of interconnected nodes—these networks have no distinguishable layers of invariant representations. With an explicit hierarchical network abstract models of collectively derived associations can be created. One such layered network can be generated for the scientific community. As scientist go about their research, a multi-layered network consisting of scientists, their papers, and their journals can be derived from digital-library metadata. The individuals, their papers, and their journals form the three layers of the hyper-cortex. Each layer interconnects its nodes with horizontal projections: authors are associated in thought by their co-authored papers, papers are connected to one another by their citations, and journals and conference proceedings are related by their domain. As one moves up the scientific community's hyper-cortical hierarchy the breadth of the invariant representations become more pronounced. Authors, as general problem-solvers of their domain, are lossy models of the specific problems facing the world—many problem's map to one author. Written papers are lossy models of the knowledge contained within their authors and referenced authors—many authors map to one paper. And finally, journals and conference proceedings are lossy models of the papers they contain—many papers map to one journal. The further up the abstraction layer, the more general the concept, the more invariant the representation, and the more information lost to the necessity of compression. It is within the domain of scientific publications that an explicit, well documented, hierarchical hyper-cortex has already begun to emerge. And it is one of the goals of this paper to extract this information to construct a collective problem-solving hyper-cortex.

## 3b. Collective Attention as the Flow of Information-Molecules

At any moment, the collective mind is thinking about a particular topic. This collectively generated topic can be broad, spanning, what seems to the individual, a confusion of unrelated ideas. In a momentary span of attention there exists a collective subjective experience of the collective's external world. The collective is currently thinking about its environment; internalizing the activation patterns of its individual's current problems to continuously transform them into associated models within and between the layers of its hyper-cortex. This continuous spread of hyper-cortical activation is the collective thought—an expression of collective problem-solving.

Millions upon millions of individuals are currently moving through the World Wide Web—looking up web pages that relate to their questions, and, in turn, finding solutions to their problems. At no point in time has the World Wide Web been viewed in its entirety by every member of the collective. With billions upon billions of web pages and only millions upon millions of users, at any one moment only a small percentage of the web is actually being viewed. The perceived subset of the web is the focus of the *collective attention*—a pattern of activation representing the populations currently problem-solving activity [12]. Collective thought then is the evolution of this pattern through time as people link between pages and use search engines to jump between disparate ideas. The individuals, or *information molecules*, through their aggregate linking, create the fluidity collective mind's experience [13]. These molecules have a goal, a history, and a method to their behavior. Their environment has a structure that constricts the absolute mobility of their movements. Each information molecule is a unique cortical problem-solving system. How do all these information molecules contribute to the collective problem-solving process? How does the interaction of all the individuals generate a solution to a problem greater than those experienced by their own individuality?

In the collective-mental map presented in [11], these information molecules habituate the hyper-cortex network through repeated selection of ingrained paths—thus restricting the future behaviors of other information molecules. This is a form of *stigmergy*. A single information molecule affects the probability of another molecules path by altering the external environment. In this system, each word-node in the network has a list of other word-nodes that an individual can link to. This list is ranked according to how many other individuals took the same path—popular paths are higher on the list. The ranked list of each node primes future individuals to choose well ingrained paths and thus restricts the probability of their choices. The World Wide Web works in a similar fashion. Web page authors alter the probability of an information molecule's next step via web links. Search engines do the same with their page ranking algorithms. In the scientific community journals have impact ratings, manuscripts have references, and digital library search engines have discipline categories. These structural biases cause the collective mind to focus on popular regions of the collective mental-map, which, in turn, further incites a realization of the world within the terms of this focus. This is the pattern-match of the current world perception to a popular invariant representation stored in the hyper-cortical memory. The continuous flow of the collective's attention within trend regions of the hyper-cortex can further increase the probability of future activity in that area of the network. Future experiences of the collective have a higher probability of being modeled by these trend representations. This is why terms such as 'complex systems', 'scale-free networks', and 'multi-agent systems' are the abstract models that the collective currently uses to represent their problems and solutions. With virtually unlimited space on the web and in the federated digital libraries, old ideas are not forgotten; they are simply not attended too. They exist structurally, but the probability of an information molecule reaching these ideas is low. Therefore, outdated invariant representations within the hyper-cortex are not tied to newer experiences. Where, on the other hand, the invariant representations of the collective's attention are the epicenter of novelty and neuronal growth within the collective mind.

### 3c. Problem Abstraction and Solution Specification

As stated before, when an individual experiences their environment, the cortex automatically categorizes the sensory pattern in relation to its invariant representations. In order to move that information outside their head and into a medium that is transparent to others within the collective, that information must be explicitly represented. Within the scientific community, this explicit representation is a written manuscript—a paper formalizing a solution to a perceived problem in the environment. This is the highest-level of abstract problem-solving that the single individual can accomplish. The individual, upon completing a paper, can store this novel solution model in hyper-cortex's paper layer. In the future, new problems of the community can potentially, after a series of transformation, return the individual's written manuscript as a potentially usable solution model.

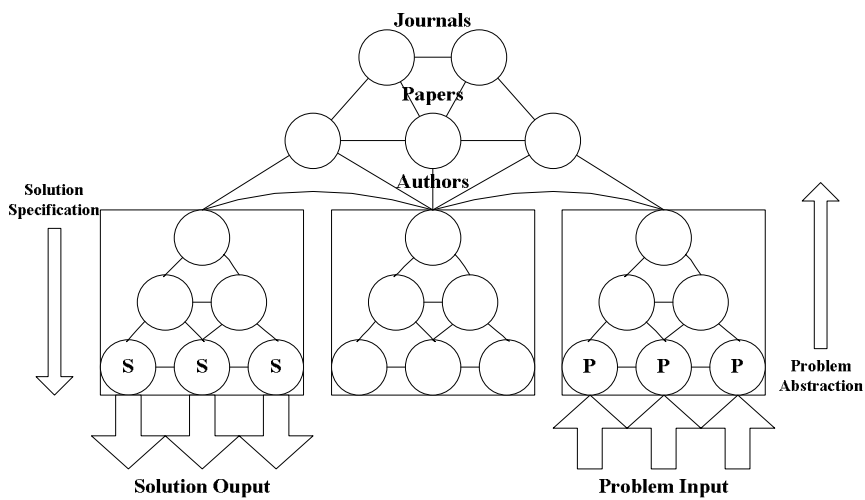

Figure 3: abstracting a problem outside the individual and implementing the solution by another

An individual perceives their world, abstracts it, and represents it within the community. The community references it, organizes it, and provides the means for another individual to use that model for finding their solutions. This model is already in use by the scientific community's current non-tangible cortical architecture. Individuals and their collaborators perceive problems in the environment and then solve them. These solutions are represented as papers. These papers are categorized in journals and conference proceedings. Other individuals within the domain subscribe to these journals. Subscribing individuals read the focused papers of these broad ranging journals. The information in these papers alters the reader's internal mental-map—causing them to make new associations between disparate ideas or simply reinforcing old ideas. These new invariant representations become the focus of their attention and therefore the means by which they perceive the new problems of their environment. And so within the human and collective cortices, the problems of one individual, through the process of abstraction and specification, become the solutions of another (Figure 3).

The scientific community's publication process has existed for many years, but only in the last 15 years has the community moved their work into the computational domain—into the digital-library domain. This explicit representation of the community provides the necessary medium to algorithmically optimize the publication process—creating a more friction-less scientific community. The next two sections will create a hyper-cortex from the metadata contained in digital-library systems and then describes the way the hyper-cortex solves five common problems in the scientific community.

## 4. Constructing the Scientific Communities Hyper-Cortex

The scientific community's hyper-cortex has been made more explicit due in large part to the extensive use of digital-libraries. Many institutions use digital-libraries to electronically publish the pre-prints of their research findings. Digital-library technology not only stores digital copies of written manuscripts, it also maintains a record of each paper's associated metadata. Manuscript metadata such as authoring scientists, cited papers, and publishing journals make it possible to algorithmically generate the scientific community's hyper-cortex. This section will describe how to utilize digital-library metadata to construct the three layers of the hyper-cortex and interconnect them into a problem-solving hierarchy.

## 4a. Digital-Library Metadata Review

The most important metadata specification in use to date is the OAI[2]-metadata standard [14]. The OAI-PMH is the associated protocol used to harvest OAI-metadata from digital-libraries [15]. Initially this standard was used exclusively for searching, but recent research has recommended the development of more advanced services [16]. Before discussing how the OAI-metadata can be used to generate a digital representation of a hyper-cortex, the OAI `<record>` container will be reviewed. A `<record>` in the OAI specification contains all the metadata for a specific repository item—a submitted paper. The randomly chosen CiteSeer [17] record below is represented within the Dublin Core (`<dc:>`) and CiteSeer (`<oai_citeseer:>`) namespaces. For the sake of brevity, multiple tags that do not pertain to the goals of this paper have been left out. Also the abstract description and the references section at the end of the record have been shortened. Important tags that will be referenced in the following subsections are highlighted with bold font and have the Δ# notation near the right hand margin.

```
<record>
  <header>
    <identifier>oai:CiteSeerPSU:99914</identifier>
    <datestamp>1998-10-05</datestamp>
  </header>
  <metadata>
  <oai_citeseer:oai_citeseer
xmlns:oai_citeseer="http://copper.ist.psu.edu/oai/oai_citeseer/" xmlns:dc
="http://purl.org/dc/elements/1.1/" xmlns:xsi="http://www.w3.org/2001/XMLSchema-instance"
xsi:schemaLocation="http://copper.ist.psu.edu/oai/oai_citeseer
http://copper.ist.psu.edu/oai/oai_citeseer.xsd ">
    <dc:title>The Asymptotics of Waiting Times Between Stationary Processes, Allowing
Distortion</dc:title>
    <oai_citeseer:author name="Amir Dembo" />                          Δ1
    <oai_citeseer:author name="Ioannis Kontoyiannis" />                Δ1
    <dc:description>this paper is to extend these asymptotic…</dc:description>
    <dc:identifier>http://citeseer.ist.psu.edu/99914.html</dc:identifier>
    <dc:source>http://www.stat.purdue.edu/people/yiannis/PAPERS/ms.ps.gz</dc:source>
    <oai_citeseer:relation type="References">                          Δ2a
      <oai_citeseer:uri>oai:CiteSeerPSU:143950</oai_citeseer:uri>
    </oai_citeseer:relation>
    <oai_citeseer:relation type="References">                          Δ2a
      <oai_citeseer:uri>oai:CiteSeerPSU:50737</oai_citeseer:uri>
    </oai_citeseer:relation>
    <oai_citeseer:relation type="Is Referenced By">                    Δ2b
      <oai_citeseer:uri>oai:CiteSeerPSU:308347</oai_citeseer:uri>
    </oai_citeseer:relation>
    <oai_citeseer:relation type="Is Referenced By">                    Δ2b
      <oai_citeseer:uri>oai:CiteSeerPSU:204535</oai_citeseer:uri>
    </oai_citeseer:relation>
    . . .
    <dc:rights>unrestricted</dc:rights>
    <dc:publisher>Annals of Applied Probability</dc:publisher>        Δ3
</oai_citeseer:oai_citeseer>
</metadata>
</record>
```

## 4b. The Three Layers of the Hyper-Cortex and their Lateral Projections

The hyper-cortex described by this paper has three layers—an author layer, a paper layer, and a journal/proceedings layer. Each layer is represented using a *co-authorship network* [18] or *co-citation*

---

*network* [19], a *citation network* [20] or co-citation network, and a *domain network*, respectively. The goal of each network layer is to relate their nodes according to some homophilic property—some quality of relatedness. For purposes of this hyper-cortex, the goal is to connect the nodes of each layer according to similar research ideas.

Co-authorship networks define the past collaborations of authors in the community. When two authors publish a paper together, an edge between them in the co-authorship network is created. It is assumed that two co-authors are similar in research ideas if they have collaborated on a project together. The strength of this relationship can be determined many ways. For one, the more two authors publish together, the greater their connection strength. If there are multiple co-authors on a particular paper, then the strength between any two co-authors is inversely proportional to the total amount of co-authors of the paper [18]. If an edge value between two authors already exists, then this new value is simply added.

$$e_{i,j} \rightarrow e_{i,j} + \frac{1}{n-1}$$

*edge weight increases according to how many co-authors (n) are co-authoring a new paper*

The lateral connections within the co-authorship network have the semantic meaning: 'has co-authored with'. An OAI `<record>` contains the necessary information to build a co-authorship network—`A1`. In terms of the paper in the previous subsection, given that there are two authors, and assuming that no other papers have been co-authored by these two authors, each directions edge weight is 1.0. Co-citation data provide another means for determining author similarity. If many papers in the field co-cite two authors, then there is a high degree of similarity between the research ideas of these two authors [19] [21]. The process to create a co-citation network is more difficult than a co-authorship network, but it is possible using both the `A1` and `A2a` entries in the OAI metadata record.

The next layer above the co-authorship layer is the paper layer. The paper layer is created using a citation network which is different than a co-citation network. A co-citation network will argue that two papers are similar if some other third paper references both of them, while a citation network states that two papers are related if either one of the two papers references the other. Directed edges in a citation network represent a reference from one paper to another. The edge strength for a particular citation link is inversely proportional to the amount of references contained in the referencing paper.

$$e_{i,j} = \frac{1}{n}$$

*edge weight is inversely proportional to the amount of reference (n) in the paper*

In a citation network, two directed edges connect the papers—feedforward and feedback. A feedforward projection states that the first paper cites the second paper—`A2a`. A feedback projection states that the second paper has been cited by the first paper—`A2b`. The two outgoing and the two incoming edges of the paper in the previous subsection each have a weight of 0.5. Co-citation is another method for constructing networks of related manuscripts. To determine author similarity by co-citation data, if two papers are cited together by some third paper, then there similarity increases.

Finally, riding atop the paper layer is the journal and conference proceedings layer, but for brevity's sake, this layer will simply be called the journal layer. The journal layer is represented by a domain network. There currently exists no research in this area and it is expected that the edges and associated strengths of the domain network can be created algorithmically by doing word analysis of papers contained within the journals or by using the user base to create a collective mental-map similar to the one described in [11]. The difficulty of defining this layer is that journals have issues and conferences have years. It may be possible to represent the journal layer into two sub-layers. The lower of the two layers represents the issue of the journal or the year of the conference, and the higher layer represents the overall journal or conference. Even with this complication, there still exist many papers in the digital-library archives that have not been published in a journal or conference proceeding—pre-prints. It is for this reason that this paper, though not discussing this idea in-depth, promotes the concept of a virtual journal [22]. The basic idea is that scientists in the community are able group electronic papers into virtual journals as they see appropriate—and store this information as a record in the digital-library archive. Scientists in the community can then subscribe to those electronic journals most related to their research. If this technology is brought to the forefront then journal similarity is determined by the amount of papers two virtual journals share in common.

$$e_{i,j} = n$$

*edge weight is equal to the amount of papers(n) the two virtual journals share in common*

| Layer | Network Type | Feedfoward Projections | Feedback Projections |
|---|---|---|---|
| Δ1-author | co-authorship | 'co-authored with this author' | 'co-authored with this author' |
| Δ2-paper | citation | 'cites this paper' | 'cited by this paper' |
| Δ3-journal | domain | 'similar to this journal' | 'similar to this journal' |

*Table 1: the semantic meanings of the feedfoward and feedback lateral projections within the hyper-cortical layers*

### 4c. The Three Layers of the Hyper-Cortex and their Vertical Projections

The previous subsection described the three layers of the scientific community's hyper-cortex and the meaning of their lateral projections. The cortical hierarchy is created by interconnecting the three layers with vertical projections. The author layer projects up the hierarchy with edges that represent that that author has written that paper. The paper layer projects down to the author layer stating that that paper has been written by that author. The paper layer then projects up to the journal layer stating that that paper has been published in that journal—Δ3. Finally, the journal layer projects down the hierarchy stating that that journal contains that paper.

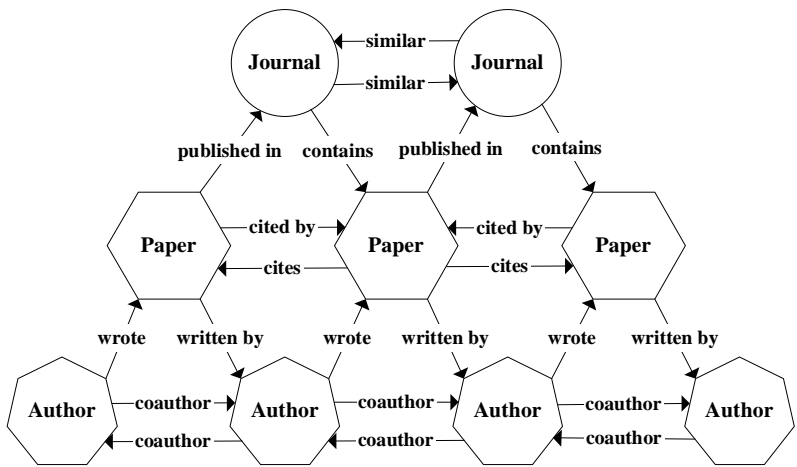

*Figure 4: digital-library metadata can create the vertical and horizontal projections of the scientific community's hyper-cortex*

In Figure 4, upward projections are on the left hand side of the object and downward projections are on the right hand side. The imbalance of projections was done to preserve the clarity of the diagram.

| Layer | Projections Down | Projections Up |
|---|---|---|
| Δ1-author | n/a | 'wrote this paper' |
| Δ2-paper | 'written by this author' | 'published in this journal' |
| Δ3-journal | 'contains this paper' | n/a |

*Table 2: the semantic meanings of the horizontal and vertical projections in the hyper-cortex*

With today's current digital-library technology, layer's 1 and 2 are easily constructed. What is not so obvious is the lateral and horizontal projections within and between layer 3. As stated before, it is the hope that the ideas in this paper will stimulate the development of a virtual journal record in the OAI-specification. In next section, a specific hyper-cortex pattern-matching algorithm is used to solve five common problems in scientific community.

# 5. Problem-Solving in the Scientific Community's Hyper-Cortex

Collective problem-solving, like in individual problem-solving, is a pattern-matching function. Given a model of the environment (the problem)—an author, a paper, a journal, or some collection of the three—the hyper-cortex will then perform a series of transformations to the problem in order to derive a collection of potential solutions. The basic idea is to continuously map the input signal within and between layers of the hyper-cortex until some desired format emerges.

Each problem of the following subsections is represented graphically as a three layered hyper-cortex. A set of input nodes, or *problem-model*, is represented as a collection of either a P+ or P- nodes. P+ nodes refer to nodes that are the initial sources of positive or excitatory energy. P- nodes are nodes that are the initial sources of negative or inhibitory energy. Depending on the context of the problem, some energy should be excitatory and others inhibitory. The output nodes, or S nodes, represent the *solution-model* and are determined at the end of the problem-solving process. Energy is contained within discrete information molecules or *particles*. Each time step, a particle takes a particular an outgoing edge of its current node depending on a probability derived by normalizing all outgoing edge's weight to 1.0. With each time step, the energy content of the node decays according to a *decay scalar* parameter—a value between 0.0 and 1.0. A 1.0 decay scalar means that at each time step 100 percent of a particle's energy content remains and therefore the total energy in the network will never decrease. A decay scalar of 0.0 means that after one time step all energy in the particles will decay to 0.0. Whenever a node receives an incoming particle it adds the energy content of that particle to its current energy content—for negative energy, the nodes total energy is decreased. It is important to remember that for each problem-solving process each node keeps a history of how much energy has passed through it. The particle dissemination algorithm continues until the energy content of all the particles has reach 0.0—until the decay scalar has completely reduced each particle's energy content. The set of all nodes which have an energy value greater than 0.0 is considered the solution-model. The solution-model can be trimmed using an *energy threshold value* if the S node set is too large. In the scientific community's hyper-cortex there can be multiple S node sets: $S_A$, $S_P$, and $S_J$. The various S sets are solution-models in author, paper, and journal format. The actual solution to the problem is dependant upon the observers desired solution format—which is dependant on the problem's context.

The following subsections take the reader through five stages of the paper writing process. The community's hyper-cortex can support a scientist through the development of an initial idea to the distribution of the idea's manuscript to interested members in the collective. Each of the five major problem-solving endeavors are possible via problem/solution pattern-matching.

## 5a. Realizing References

It is the goal of any scientist to publish papers. A scientist must use their internal cognitive faculties and the past and present ideas of their colleagues to generate novel solutions to the problems facing their discipline. Every paper starts with some initial idea that grows into a formalized model ready for reading by their colleagues. In many cases that idea is triggered by some paper they read. From here, this novel idea must be nurtured by and tied to other papers in the domain. So, from an initial idea, sparked by some keystone paper, the scientist must find the niche for this new, soon to be formalized, idea. The keystone paper of the paper currently being read was the book by Hawkins and Blakeslee [1]. From there many papers from the discipline of neuroscience, collective-intelligence research, and digital-library technology were sought to give the paper context within the community.

For this individual scientist's problem, the $P_{P+}$ set represents the keystone paper. This is the only sensory information the hyper-cortex has and therefore this small set is the hyper-cortex's problem-model. In order to trigger the cascade of energy through the hyper-cortex, the $P_{P+}$ paper is given a collection of positive energy particles. These particles propagate within the paper layer and between the author and journal layers incrementing the energy of all the nodes they pass through. The process continues until all the energy of all the particles has decayed to 0.0. At the end of the particle dissemination process there is an $S_A$, $S_P$, and an $S_J$ set. Since the scientist is concerned with finding potential references for the new idea, only the $S_P$ set is reviewed. If too many papers are contained in $S_P$, the scientist can increase the energy threshold value until only the amount of papers desired is returned. For the example in Figure 5, the $S_P$ set is reduced to two nodes.

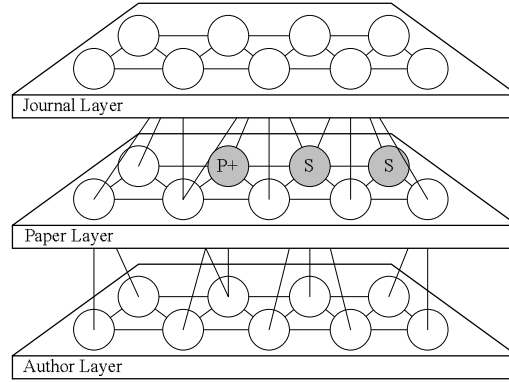

*Figure 5: $P_{P+}$ (keystone paper) → $S_P$ (related papers)*

At this point a scientist has initially found a paper that has stimulated an unformulated novel idea in their mind. The scientist has used that paper as a source for finding other related papers to further his understanding of the problem domain. This process of finding related papers is not a new idea. The hyper-cortex has always existed in the world—being partially created as individuals navigate the web of relations between authors, documents, and journals. The difference is that now the process is algorithmically performed on an explicitly represented hyper-cortex. If no explicit hyper-cortex existed, then the scientist would follow the references of the keystone paper. Review the journals that these papers are published in. Look through the curriculum vitae of the keystone paper's author for other papers. So on and so forth. This would have been a manual process, but because of the collectively generated hyper-cortex, this process can be expressed outside the individual. The scientist utilizes this collective mental-map to match his problem—find papers related to the keystone paper—to his solution—a collection of related papers worth reading.

**5b. Realizing Collaborators**

Now that the scientist has a collection of papers that are representative of his initial idea he may be interested in finding potential collaborators. A way to make this possible via the hyper-cortex is to find a collection of authors that are related to that collection of papers. The intuitive notion of this idea is that individuals whom are related in thought to the papers derived in subsection 5a are more than likely going to understand the author's vague notion of an idea and, in turn, be able to provide assistance in formalizing the idea more clearly. To determine collaborators, the input set of the hyper-cortex, $P_{P+}$, is the solution set, $S_A$, from subsection 5a.

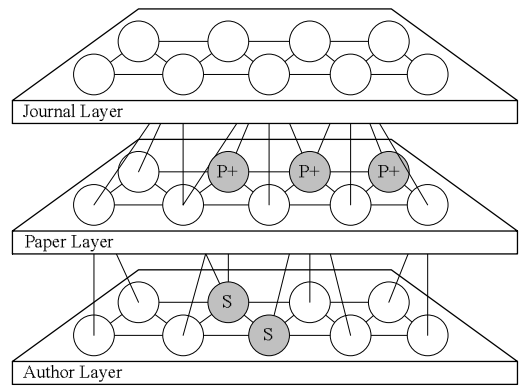

These positive energy particles propagate through the hyper-cortex—up to the journal layer and down to the author layer. Since the scientist is interested in finding collaborating authors the scientist reviews the $S_A$ solution-model once the energy content of all the particles has decayed to 0.0. The collection of authors in $S_A$ can then be contacted for discussion about these ideas and a potential collaborative paper writing relationship may form.

## 5c. Realizing Journals to Publish In

After finding papers to reference and other scientists to discuss the initial vague idea with, the original scientists and his collaborators have written a paper that contains a novel idea that they feel is worth publishing in a journal. In order to find a good arena to publish their work in, the authors must find a journal that is related to the now formalized paper. There could be many ways to organize this problem-model, but for the example demonstrated in Figure 7, the problem-model is the set of nodes in both the paper and author layer. $P_A+$ is the set of all collaborating authors on the new paper. Next, in the paper layer, the $P_P+$ set is the set of all papers referenced by the new unpublished paper. After the initial distribution of all positive energy, the collective thought process runs until the energy content of all the particles has decayed to 0.0.

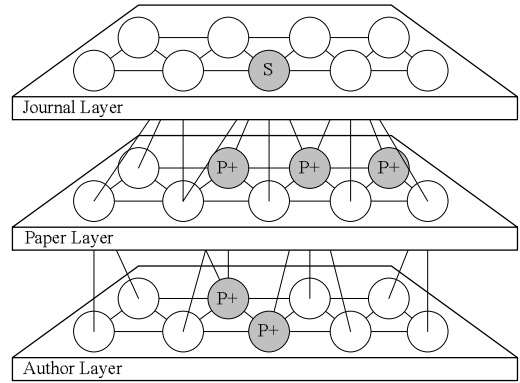

*Figure 7: P_P+ (referenced papers) · P_A+ (collaborators) → S_J (submitting journal)*

Since the scientists are interested in a solution that refers to a potential journal to publish in, the scientists set the energy threshold value to the value of the highest energy node in the journal layer. This trims the $S_J$ set to one node—the journal to submit the paper to. The intuitive idea is to search the hyper-cortex for journals that have published the papers of the cited articles or of individuals whom are related in thought to the collaborating scientists.

## 5d. Realizing Peer-Reviewers

With the paper written and submitted it is up to the journal to determine the quality of the work. This is accomplished by means of the peer-review process. The editors of the journal locate individuals within the community for whom they believe to be experts in the paper's domain. The paper is then distributed to these referees for review. Sometimes the paper is accepted as is, other times it's returned for revision, and still other times, the paper is rejected. It is up to the reviewers to determine whether the paper is deemed of high enough quality to allow others in the community to use the work in their future publications. Finding experts in the field can be a laborious task of manual effort. The hyper-cortex can solve this problem by matching an unpublished paper to a set of expert individuals in the author layer. The unpublished paper is abstractly represented by the $P_P+$ set as all its references since it's representative node hasn't been placed in the hyper-cortex—since no invariant representation of the paper exists. Secondly, a $P_J+$ set can be created with one node—the reviewing journal. The reason for stimulating the reviewing journal is because the reviewing journal may want to search the web of authors whom have published in their journal before knowing that they understand the quality requirements of the journal. Finally, and of most interest is the $P_{A-}$ set. This is the set of authors whom have written the paper. Since an author cannot be a referee to their own paper, and since it is advisable that past collaborators of those authors also not be referees, these nodes should be given negative energy to inhibit the activation of themselves and the individuals around them in the co-authorship network.

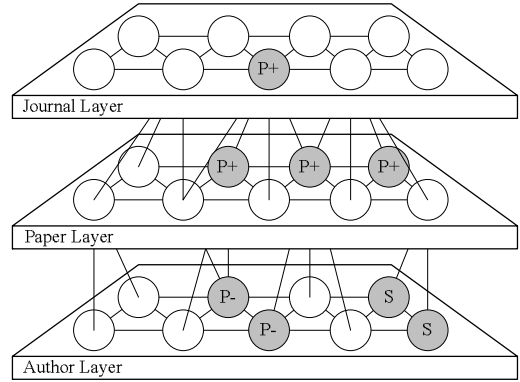

*Figure 8: $P_{J+}$ (reviewing journal) · $P_{P+}$ (referenced papers) · $P_{A-}$ (collaborators) → $S_A$ (reviewers)*

For an in depth look at this process please refer to [23]. In [23] the algorithm was carried out solely in the author layer by utilizing the authors of the referenced paper (instead of the referenced papers themselves) as the problem-model. Furthermore, what is of interest is that the set $S_A$ can have an energy threshold value of 0.0. By doing this, every individual in the output set $S_A$ is preserved. Each member in that set then has some proportion of the total network energy relative to the others in the collective. This varying degree of energy can be used to determine the individual's decision-making influence in the review process—where more expert individuals, with regards to the paper in question, should have more say as to the quality of the paper. This idea has been proposed as a means of augmenting digital-library technology to support the peer-review process and is further developed in [23].

## 5e. Realizing Readers

Finally the paper has been written, submitted, and accepted. It is the role of the journal to find the largest distribution base for the paper. Journals, that are not fee-based, can aid the author and the community by matching the paper to a set of readers who may be interested in the newly published paper. This is done by creating a $P_{P+}$ set of all the referenced papers of the newly published paper and creating a $P_{A+}$ set of the authors of the paper. The particle dissemination algorithm runs and the solution set, $S_A$, is the set of all individuals in the author layer whom received an amount of energy greater than 0.0. If these individuals have provided email address with the digital-library system, then these authors can be automatically contacted with an email containing the title of the paper, its abstract, and a link to it in the digital-library repository. Authors can increase the energy threshold value of their author node if they wish to receive only those papers that are most pertinent to them.

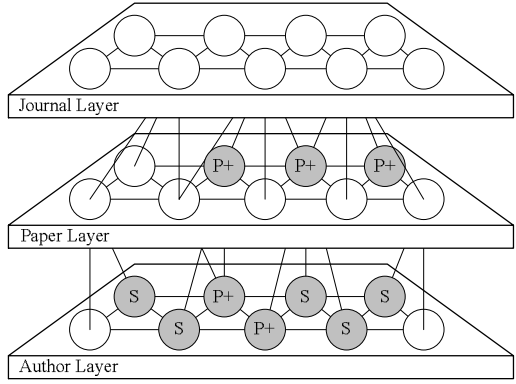

*Figure 9: $P_A+$ (paper authors) · $P_P+$ (referenced papers) → $S_A$ (interested readers)*

| P/S process | Problem-Model | Solution-Model |
|---|---|---|
| References | $P_P+$ (related keystone paper) | $S_P$ (related papers) |
| Collaborators | $P_P+$ (related papers) | $S_A$ (potential collaborators) |
| Journals | $P_P+$ (related papers) · $P_A+$ (co-authors) | $S_J$ (submitting journal) |
| Peer-reviewers | $P_J+$ (journal) · $P_P+$ (related papers) · $P_A-$ (co-authors) | $S_A$ (potential reviewers) |
| Readers | $P_P+$ (related papers) · $P_A+$ (co-authors) | $S_A$ (interested readers) |

*Table 3: the problem-model and solution-models of 5 hyper-cortical problem-solving processes*

A single hyper-cortical structure can provide the functionality to perform many complex problem-solving tasks for the community (Table 3). By testing various parameters such as the amount of particles to distribute, the energy threshold values, and the multiplicative properties of particle interaction, these basic ideas can be finely tuned to meet the needs of the user. There are still many other problem's that can be developed that have yet to be realized by this paper. With future research into the hyper-cortex architecture and its particle-flow problem-solving process, digital-library's can be augmented above and beyond simple manuscript searching.

## 6. Conclusion

What has emerged in the scientific community is computational medium that supports collective problem-solving. Each individual stigmergetically contributes to the evolution of this structure and therefore the potential paths of its users. The collective brain continues to grow at a phenomenal rate as more and more individuals publish their work into the federated OAI digital-libraries. How will such a hyper-cortex affect the collective-attention? How does the hyper-cortex promote the self-organization of the relationships of the community members and their work? A well engineered hyper-cortex is the foundation for a 'friction-less' community [24]. What this means is that information will get to where it needs to go when it needs to get there—finely tuning the community by meditating its information flow and member relations. This is in line with the goals of the Open Archives Initiative. A hyper-cortically supported scientific community is a self-organizing entity that constantly derives solutions to its problems by matching its present state with its past realizations via the use of its artificial neural-network.

This paper has provided a glimpse into a collective-intelligence paradigm extended from work done within the individual-intelligence research domain. Other such hyper-cortical structures can be created. A movement away from flat networks is needed if more abstract problem-solving is to be possible. Other collectively generated social-networks can make use of this research to facilitate a new form of human-collective problem-solving unseen in today's current technology.

## 7. Acknowledgements

This paper's production is due in large part to funding from the U.S. Department of Education's GAANN Fellowship and the Fonds voor Wetenschappelijk Onderzoek - Vlaanderen. A special thanks goes out to Francis Heylighen and Carlos Gershenson for extended talks on this topics.